\documentclass[manuscript]{aastex}

\usepackage{emulateapj5}
\submitted{Accepted for publication in the Astronomical Journal}

\received{}
\accepted{}

\newcommand {\apgt} {\ {\raise-.5ex\hbox{$\buildrel>\over\sim$}}\ }
\newcommand {\aplt} {\ {\raise-.5ex\hbox{$\buildrel<\over\sim$}}\ }

\begin{document}

\title{The Starburst Nature of Lyman-Break Galaxies:\\ Testing UV
Extinction with X-rays}

\author{Mark Seibert, Timothy M. Heckman and Gerhardt R. Meurer}

\affil{The Johns Hopkins University, Department of Physics and Astronomy\\
3400 North Charles Street, Baltimore, MD 21218}
\email{{mseibert@pha.jhu.edu}, {heckman@pha.jhu.edu}, {meurer@pha.jhu.edu}}

\shorttitle{Testing the UV Extinction of Lyman-Break Galaxies} 
\shortauthors{Seibert, Heckman, \& Meurer}
 
\begin{abstract}
We derive the bolometric to X-ray correlation for a local sample of
normal and starburst galaxies and use it, in combination with several UV
reddening schemes, to predict the 2--8 keV X-ray luminosity for a sample
of 24 Lyman-break galaxies in the HDF/CDF-N. We find that the mean X-ray
luminosity, as predicted from the Meurer UV reddening relation for
starburst galaxies, agrees extremely well with the Brandt {\it stacking}
analysis. This provides additional evidence that Lyman-break galaxies
can be considered as scaled-up local starbursts and that the locally
derived starburst UV reddening relation may be a reasonable tool for
estimating the UV extinction at high redshift. Our analysis shows that
the Lyman-break sample can not have far-IR to far-UV flux ratios similar
to nearby ULIGs, as this would predict a mean X-ray luminosity 100 times
larger than observed, as well as far-IR luminosities large enough to be
detected in the sub-mm.  We calculate the UV reddening expected from the
Calzetti effective starburst attenuation curve and the radiative
transfer models of Witt \& Gordon for low metallicity dust in a shell
geometry with homogeneous or clumpy dust distributions and find that all
are consistent with the observed X-ray emission. Finally, we show that
the mean X-ray luminosity of the sample would be under predicted by a
factor of 6 if the the far-UV is unattenuated by dust.
\end{abstract}


\section{Introduction}

Lyman-break galaxies (LBGs) represent the sites of a significant
fraction of the star formation in the early universe (e.g., Adelberger
\& Steidel 2000) and the ultraviolet (UV) derived star formation rates
(SFRs) of LBGs have become an important component of the recent attempts
to measure the star formation history of the universe (e.g., Madau et
al. 1996; Steidel et al. 1999; Lanzetta et al. 2002). Precisely how the
UV emission of LBGs are corrected for the effects of intrinsic dust
attenuation, which requires a knowledge about the true nature of these
objects, is critical for estimating star formation rate densities at
$z>2$.

Starburst galaxies are the most likely local analogs to the population
of LBGs at high redshift.  LBGs have UV surface brightnesses and UV
colors similar to local starbursts (Meurer et al. 1997) suggesting that
LBGs can be considered as larger and more luminous versions of the
nearby variety.  Further evidence in support of the analogy comes from
the only high quality rest-frame UV spectrum of an individual LBG -- the
gravitationally lensed MS 1512+36-cB58 -- which shows emission and
absorption features remarkably consistent with local starbursts (Pettini
et al. 2000).

Local starburst galaxies exhibit an empirical correlation between their
UV spectral slope ($\beta$; equivalent to UV color) and their infrared
excess (IRX; ratio of far-IR to far-UV flux). This relationship implies
that starbursts redden as dust absorption increases and that dust
surrounding starburst regions can be modeled to provide an estimate of
the UV extinction as a simple linear function of the UV color (Meurer et
al. 1999; hereafter MHC99). Although the UV properties of LBGs can be
readily measured from deep optical imaging, obtaining rest-frame far-IR
data for these high redshift galaxies is extraordinarily
difficult. Therefore, it is unclear whether or not the IRX-$\beta$
reddening relationship is a meaningful tool that may be used to estimate
the UV extinction of high redshift galaxies. Until rest-frame far-IR
data become available for a significant sample of LBGs, we must attempt
to probe other accessible wavelength windows that may provide clues
about the thermal dust emission.  X-rays can be just such a proxy.

In this paper, we investigate the suggested starburst nature of LBGs by
considering their X-ray emission and the consequences for UV
extinction. To do this, we first use a local sample of normal and
starburst galaxies to demonstrate a clear correlation between the
bolometric and 2--8 keV X-ray luminosities when the bolometric output is
dominated by far-IR and far-UV emission. We then test the starburst
concept for LBGs by applying the starburst derived IRX-$\beta$
relationship to a sample of 24 LBGs in order to predict their X-ray
emission. Specifically, we estimate the far-IR luminosity using the
starburst UV reddening relationship and then estimate the X-ray
luminosity from the bolometric to X-ray correlation. These results are
compared to the recent 1 Ms exposure Chandra Deep Field North (CDF-N)
X-ray constraints of Brandt et al. (2001) for the same sample.

It is possible that the UV properties of LBGs are only superficially
starburst-like. Perhaps they are the UV-bright portion of `scaled-up'
ultraluminous infrared galaxies (ULIGs) and the majority of UV light
generated by star formation is heavily extincted and not detected in
rest-UV surveys (e.g., Goldader et al. 2002). Certainly the irregular
morphology of LBGs is consistent with the notion that they are recent
mergers or interacting systems. Alternatively, LBGs may be youthful
proto-galaxies in the process of forming the spheroids destined to
become the central bulges of spirals or elliptical galaxies. If so, LBGs
could be relatively free from the effects of dust because they have not
yet been enriched with a significant quantity of heavy elements. These
scenarios represent the extremes of UV extinction in LBGs and,
consequently, extremes of their estimated star formation rates (SFRs) as
well. We test several alternative hypotheses by adapting our technique
to estimate the X-ray luminosities assuming that LBGs 1) have far-UV and
far-IR properties consistent with those found by Goldader et al. (2002)
for ULIGs, 2) have low metallicity (SMC-type) dust as modeled by Witt \&
Gordon (2000) or parameterized by the effective attenuation curve for
starbursts (Calzetti et al. 2000) or 3) have stellar populations
unattenuated by dust.

\section{The Lyman-break Sample}

We adopt the sample of 24 Lyman-break galaxies from the Hubble Deep
Field North (HDF-N) survey region within the CDF-N used by Brandt et
al. (2001), so that we may compare our X-ray predictions to their X-ray
measurements. The sample selection process we follow is described in
detail by Brandt et al., and is based upon two criteria; 1) galaxies are
spectroscopically confirmed as high redshift with $z=2$--$4$ and 2) none
is individually detected or are within $\sim8$\arcsec\ of a detected
source by {\it Chandra}. The spectroscopic redshifts are compiled from
the Caltech Faint Galaxy Redshift Survey (Cohen et al. 2000) and the
serendipitous catalog of Dawson et al. (2001). The sample is listed in
Table~1. We use the observed-frame optical measurements of the HDF-N
catalog (Williams et al. 1996). The mean (median) redshift of the sample
is 2.79 (2.93).

\section{The Observed X-ray Emission of the Lyman-Break Galaxy Sample}

The observed mean X-ray emission for the LBG sample has recently been
determined by Brandt et al. (2001) by {\it stacking} the counts of each
galaxy from the 1 Ms CDF-N survey data into an image with an effective
exposure time of 260 days (22.4 Ms). They found 43 counts within the 30
pixel aperture of the {\it stacked} image for the soft (0.5--2 keV
observed-frame) bandpass. The expected background counts were 26.2. The
statistically significant detection (99.9\% confidence level for Poisson
statistics) gives an average count rate of $\,7.5 \times 10^{-7}\
$count$\,$s$^{-1}$. The {\it Chandra} soft band corresponds to a
rest-frame bandpass of 2--8 keV at the median redshift of the
sample. The opacity of interstellar gas in this spectral range is
comparable to the opacity at $\lambda=3$--$60\,\micron$ (Morrison \&
McCammon 1983; Draine \& Lee 1984). For a neutral interstellar medium
with a column density of $N_{\rm H}=1.0 \times 10^{21}$ cm$^{-2}$ the
optical depth at 2 keV ($\tau_{\rm 2keV}$) is only 0.04 (Morrison \&
McCammon 1983).

Brandt et al. assume an intrinsic power law spectrum with photon index
$\Gamma = 2$ (typical of X-ray binaries and low luminosity AGN) and a
mean column density for the CDF-N of $N_{\rm H}=1.6 \times 10^{20}$
cm$^{-2}$ in order to estimate a mean flux of $4.0 \times 10^{-18}\
$erg$\,$s$^{-1}\,$cm$^{-2}$. The mean observed rest-frame 2--8 keV
luminosity is $2.8 \times 10^{41}\,$erg$\,$s$^{-1}$ for $H_0=70$
km$\,$s$^{-1}\,$Mpc$^{-1}$ ($\Omega_M=1/3$, $\Omega_{\Lambda}=2/3$) at
the sample's median redshift. 

We crudely estimate two sources of possible error for the mean flux: 1)
49\% random error from photon counting and 2) 30\% systematic error due
to choice of intrinsic X-ray spectrum. The photon counting error is
taken as the quadrature sum of the total count error ($\pm\sqrt{43}$)
and the background count error ($\pm\sqrt{26.2}$). The systematic
error is assessed by computing the differences in X-ray flux assuming
variations in $N_{\rm H}$ and $\Gamma$ for the power law spectral model
as well as alternative Raymond-Smith plasma and thermal bremsstrahlung
models. Combining the random and systematic errors in quadrature yields
a total error estimate of $\sim58$\% (log L$_{\rm X,obs} = 41.45 \pm 0.25$).

\section{The Local Bolometric to X-ray Correlation}

\begin{figure*}[th]
\epsscale{1.5}
\plotone{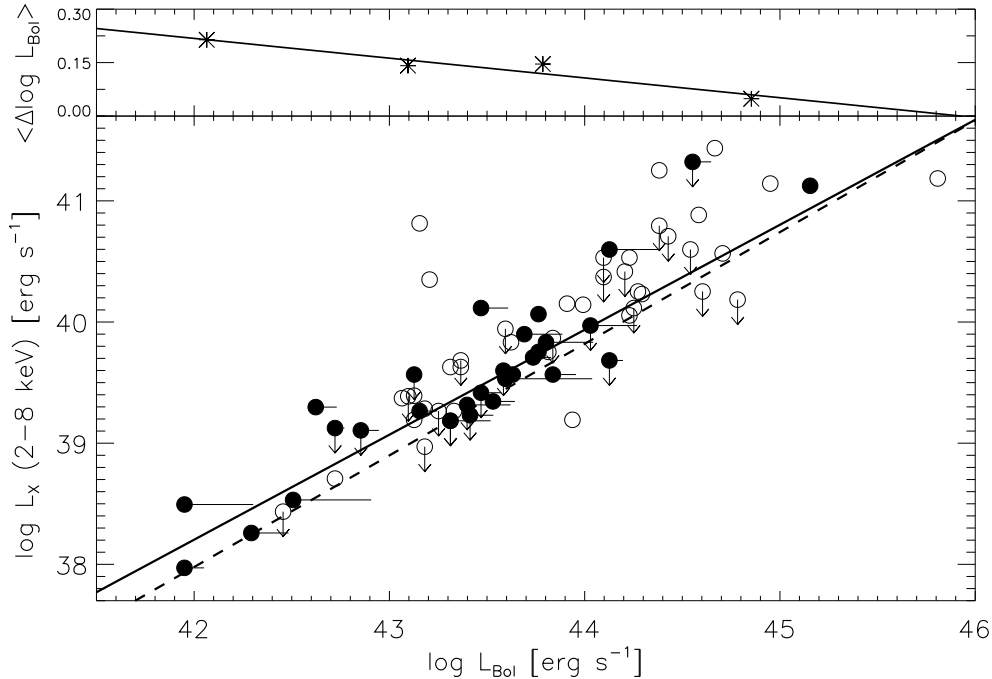}
\figcaption[fig1.eps]{BOTTOM: The bolometric to X-ray correlation for
normal and starburst galaxies. The sample is from David et
al. (1992). The X-ray luminosity is derived from {\it Einstein} IPC
data and corrected for bandpass and spectral model. All circles
represent L$_{\rm Bol}$ estimated from the far-IR only (L$_{\rm
Bol[FIR]}$). Solid circles represent a subsample where far-UV data is
also available and the effect on L$_{\rm Bol}$ is shown with
horizontal lines (L$_{\rm Bol[FIR,FUV]}$). The solid line is the fit
to the entire sample (71) using L$_{\rm Bol[FIR]}$.  The dashed line
is the fit to the subsample (29) using L$_{\rm Bol[FIR,FUV]}$. The
fits account for data points with the $3\sigma$ X-ray upper limits
indicated with downward arrows (see text). TOP: The difference between
log L$_{\rm Bol[FIR,FUV]}$ and log L$_{\rm Bol[FIR]}$ for the
subsample. The data has been grouped in bins of 1 dex and the mean
difference is plotted against mean log L$_{\rm Bol[FIR]}$. The line is
a simple chi-square linear fit.
\label{fig1}}
\end{figure*}

Several authors have published far-IR to X-ray correlations for a wide
variety of galaxy types and AGNs (e.g., David et al. 1992; Green et
al. 1992; Ranalli et al. 2002). We have adopted the normal and starburst
galaxy sample of David et al. (1992) as our starting point for defining
the bolometric to X-ray correlation because of the large sample size
(71) and the consistent spectral model employed to obtain X-ray fluxes
from count rates. The X-ray data used by David et al. are from {\it
Einstein} IPC (0.5--4.5 keV) observations and fluxes are derived
assuming the intrinsic emission is given by a $kT=5$ keV Raymond-Smith
plasma with solar abundances and a mean hydrogen column density of
$N_{\rm H}=10^{20}$ cm$^{-2}$. Far-IR fluxes are based on IRAS 60 and
100 \micron\, flux densities as defined by Helou et al. (1988).  David
et al. computed X-ray and far-IR luminosities for the sample using
distances from Sandage \& Tammann (1975, 1981) or from the redshifts
published by Soifer et al. (1987). We adjust all luminosities to reflect
our adopted $H_0$.

As our intention is to predict the rest 2--8 keV X-ray luminosity of the
Lyman-break sample we apply a correction to the David et al. X-ray flux
measurements that will account for the difference in bandpass and
intrinsic spectral model. Using the X-ray spectral fitting package XSPEC
(Arnaud 1996) we have modeled the spectrum in the 0.5-8 keV range as a
power law spectrum with photon index $\Gamma = 2$. The photon count rate
in the 0.5--4.5 keV range was normalized to that expected for a 0.5--4.5
keV flux of $5.5 \times 10^{-18}\ $erg$\,$s$^{-1}\,$cm$^{-2}$ (the mean
of the David et al. sample) from a 5 keV Raymond-Smith plasma with solar
abundances. Photoelectric absorption was applied using $N_{\rm
H}=10^{20}$ cm$^{-2}$. The result is a correction factor of order unity
(F$_{2-8,{\rm PL}}$/F$_{0.5-4.5,{\rm RS}}=0.6$).

We estimate the bolometric output of this star forming galaxy sample by
assuming that it is dominated by young stellar populations. This allows
us to approximate the bolometric luminosity from the far-IR and far-UV
using;
\begin{equation}
\rm L_{Bol}=L_{FUV}\times BC_{stars} + L_{FIR}\times BC_{dust},
\end{equation}
where BC$_{\rm dust}$ is the bolometric correction for the FIR to the
1--1000 \micron\, dust emission and BC$_{\rm stars}$ is the bolometric
correction for the far-UV to the 0.0912--1 \micron\, unextincted stellar
emission of an active star forming population. We use the value of
BC$_{\rm dust}=1.75$ determined by Calzetti et al. (2000) from {\it ISO}
and {\it IRAS} photometry of a sample of eight low redshift star forming
galaxies.  We take BC$_{\rm stars} = 1.66$ ($\lambda=1600$\AA) computed
by MHC99 from a range of Starburst99 synthetic starburst spectra
(Leitherer et al. 1999). We have obtained far-UV fluxes for 29 of the 71
galaxies in the sample from either the UV catalog of Marcum et
al. (2001) at 1550\AA\ or the homogenized UV catalog of Rifatto et
al. (1995) at 1650\AA.

Of the 71 galaxies in this sample, 28 have $3\sigma$ upper limits to the
X-ray flux. We account for this fact and compute the fit to the
correlation by using the Buckley-James\footnote{From the Astronomy
Survival Analysis (ASURV) task within the STSDAS IRAF package.}
linear regression technique which computes coefficients based on
Kaplan-Meier residuals. If we consider only the far-IR emission when
approximating the bolometric luminosity (L$_{\rm Bol}$ = L$_{\rm FIR} \times$
BC$_{\rm dust}$) we can use all 71 galaxies to find the fit,
\begin{equation}
\rm log\,L_{X(2-8)} = (0.87 \pm 0.08)\,log(L_{Bol[FIR]}) + 1.81,
\end{equation}
shown by the solid line in the bottom panel of Figure~1. The error is
$1\sigma$. If we fit only the 29 with both far-IR and far-UV data (of
which 11 have X-ray upper limits) using Equation~1 we compute the fit,
\begin{equation}
\rm log\,L_{X(2-8)} = (0.92 \pm 0.10)\,log(L_{Bol[FIR,FUV]}) - 0.72,
\end{equation}
which is represented by the dashed line in the same figure.

The fit becomes slightly steeper when the UV data are properly
considered. This is due to the fact that the far-UV contributes
measurably to L$_{\rm Bol}$ in systems with lower far-IR luminosities,
but has a negligible effect on L$_{\rm Bol}$ at higher far-IR
luminosities. The effect is highlighted in the top panel of figure~1
where we plot the difference log L$_{\rm Bol[FIR,FUV]}$ -- log L$_{\rm
Bol[FIR]}$. The 29 data points have been grouped in bins of 1 dex and
the mean difference is plotted against log L$_{\rm Bol[FIR]}$. The line
is a simple chi-square linear fit. The trend is understood most simply
as a extinction effect, where the far-UV extinction is positively
correlated with L$_{\rm Bol}$ (Heckman et al. 1998). We use the the
correlation derived from the far-UV and far-IR (Equation~3) for
predicting the LBG X-ray output. It is interesting to note that the mean
observed X-ray luminosity of the LBG sample suggests a large mean L$_{\rm
Bol}$ ($\approx$ L$_{\rm FIR}$) of $\sim 10^{12}$ L$_{\odot}$. This is
at the high luminosity end of the local starburst population.

\section{X-ray Predictions}

\subsection{Case I: Estimating X-rays Assuming Starburst Reddening}

\begin{figure*}[th]
\epsscale{1.5}
\plotone{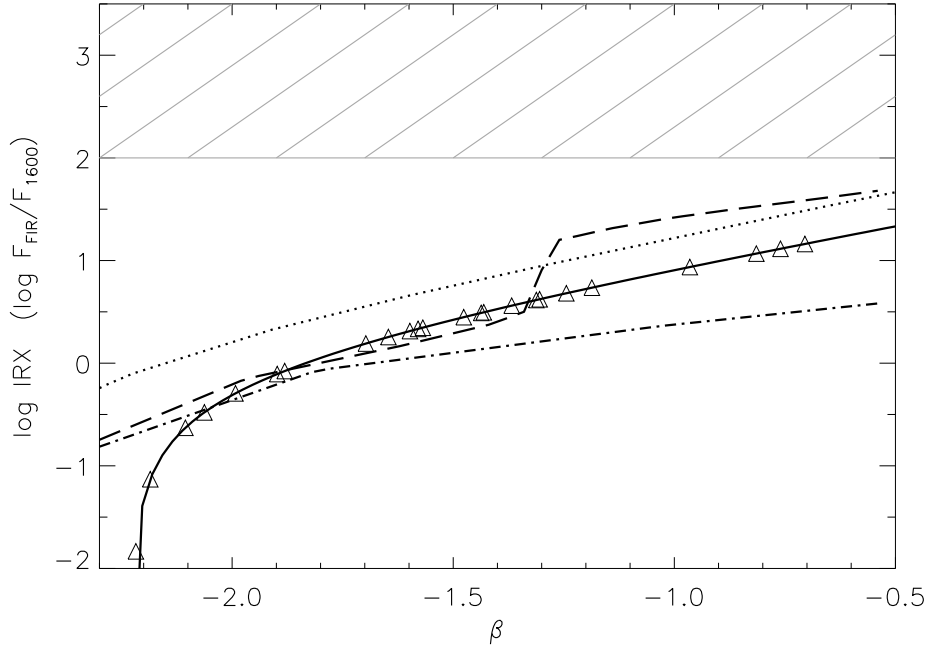}
\figcaption[fig2.eps]{UV reddening relationships used to estimate the
far-IR of the LBG sample. The solid line is the empirical IRX-$\beta$
starburst relation of MHC99. Triangles are the LBG sample with IRX
derived from the MHC99 relation and photometrically measured
$\beta$. The dash-dot line is the UV reddening relation computed from
Witt \& Gordon (2000) extinction models for SMC-dust in a shell
geometry with a homogeneous distribution. The dashed line represents a
clumpy dust distribution. The dotted line is based on the Calzetti
(2000) mean starburst attenuation curve. The hash marks indicate the
region of the IRX-$\beta$ diagram where objects with ULIG-like UV
extinction properties would be located (Goldader et al. 2002).
\label{fig2}}
\end{figure*}

The assumption that we can model the far-IR to far-UV flux ratio as a
function of UV reddening is the crucial component in our technique of
predicting the X-ray emission of the LBG sample. We first consider the
empirically derived UV reddening relation of MHC99 (IRX-$\beta$) for
local starbursts.

The UV spectral slope ($\beta$) is found from the photometric $V_{\rm
606}-I_{\rm 814}$ color. We use equation~14 of MHC99 which calibrates
the spectroscopically defined $\beta$ to the $(V_{\rm 606}-I_{\rm
814})_{\rm AB}$ color as a function of redshift between $z=$ 2--4.  The
ABmag system colors are from the isophotal fluxes in the HDF-N catalog
of Williams et al. (1996) corrected for Galactic extinction. The
IRX-$\beta$ reddening relation is then used to predict the far-IR to
far-UV flux ratio using the approximation,
\begin{equation}
\rm IRX \equiv \frac{F_{FIR}}{F_{1600}} \approx (10^{0.4A_{1600}}
-1)\times\frac{BC_{stars}}{BC_{dust}}.
\end{equation}
MHC99 demonstrated that, for starbursts, this is a reasonable estimate
for IRX and the extinction term is a simple linear function of $\beta$
(A$_{1600} = 4.43 + 1.99\beta$).

The form of the MHC99 IRX-$\beta$ relation can be seen as the solid line
in Figure~2. We estimate F$_{1600}$ by interpolating the HDF rest-frame
UV data. Equation 4 allows a prediction of the far-IR flux under the
assumption that the LBG sample has UV extinction and reddening
properties similar to local starbursts. The far-IR and far-UV fluxes are
converted to luminosities using the spectroscopically derived
redshifts. Finally, we predict the 2--8 keV X-ray luminosities for each
Lyman-break galaxy from the bolometric to X-ray correlation.

\begin{figure*}[th]
\epsscale{1.5}
\plotone{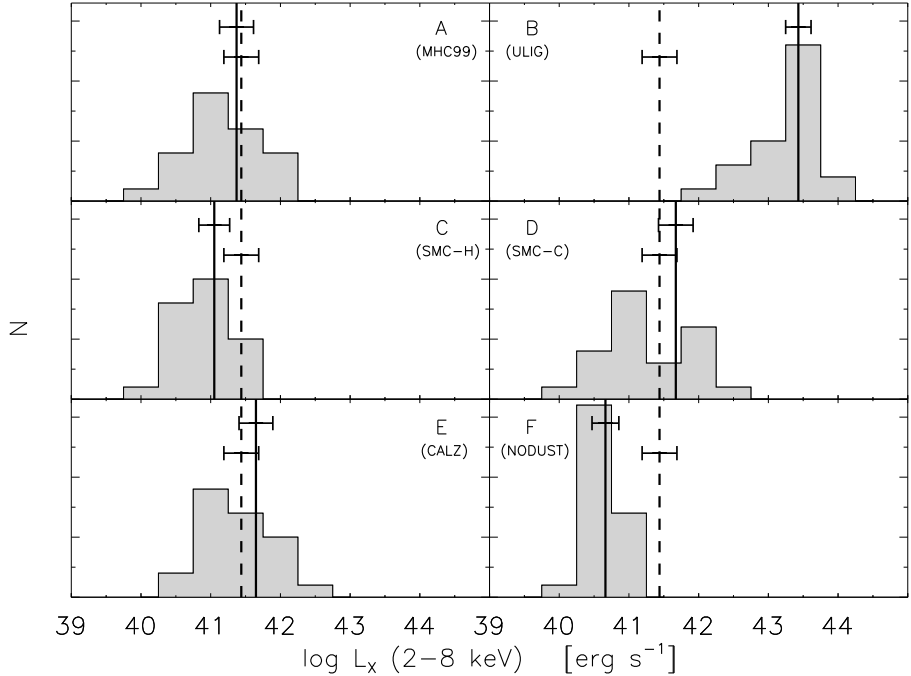}
\figcaption[fig3.eps]{Histograms of predicted 2--8 keV X-ray
luminosities for the LBG sample. The solid line is the mean of the
distribution with Monte Carlo $1\sigma$ errors. The dashed line is the
observed mean value from Brandt et al. (2001). [A] Assumes the sample
obeys the MHC99 IRX-$\beta$ reddening relationship for starbursts. [B]
Assumes UV extinction properties similar to ULIGs. [C] Assumes UV
extinction properties consistent with SMC-type dust extinction in a
shell geometry with a homogeneous dust distribution. [D] Same as panel C
but with a clumpy dust distribution. [E] Assumes UV extinction
consistent with the Calzetti starbust effective attenuation curve. [F]
Assumes the sample is unattenuated by dust.
\label{fig3}}
\end{figure*}

The $1\sigma$ error of the predicted mean is assessed with Monte Carlo
error analysis by generating 150 synthetic data sets (3600 total data
points). In addition to incorporating the measurement errors of the
HDF-N catalog into the synthetic data, the dispersion of the IRX-$\beta$
(0.25 dex in log IRX) and bolometric to X-ray correlations (0.4 dex in
log L$_{\rm X}$) are also included. The mean of each synthetic data set
is measured and the $1\sigma$ error (in dex) is taken as the standard
deviation of the distribution of differences from the true sample mean
using
\begin{equation}
{\rm Error} \; =\;\left ( \frac{1}{N}\,\sum_{i=1}^{N}\,(
  {\rm log}\,\overline{{\rm L}_{{\rm X},t}} - {\rm log}\,\overline{{\rm
  L}_{{\rm X},i}})^2 \right )^{1/2},
\end{equation}
where the subscripts $t$ and $i$ indicate the $t$rue LBG sample and
$i$th synthetic data set respectively.

The results are expressed as a histogram of the predicted X-ray
luminosities in Figure~3 (Panel~A). The log of the mean {\it linear}
luminosity is shown as a solid line with error bars. For comparison, the
mean observed value of Brandt et al. is represented as a dashed
line. The mean predicted X-ray emission is $2.4 \times
10^{41}\,$erg$\,$s$^{-1}$. This agrees well with the mean observed
value. The $1\sigma$ error is 0.22 dex. Under this scenario, all of the
LBGs in the sample are predicted to have fluxes below the single source
soft-band flux limit ($3.0 \times 10^{-17}\,$erg$\,$s$^{-1}\,$cm$^{-2}$)
for the CDF-N (Brandt et al. 2001b).

\subsection{Case II: Estimating  X-rays Assuming ULIG Properties}

Ultraluminous infrared galaxies (ULIGs) do not obey the the IRX-$\beta$
reddening relation (Goldader et al. 2002). Instead, they occupy a
distinct range of log IRX values between $\sim$2--4. Using high
resolution {\it HST} imaging data, Goldader et al. argue that the lack
of a significant IRX-$\beta$ correlation results from the fact that the
majority of far-UV light detected from ULIGs originates from the least
extincted sources (possibly just the outermost regions of star
formation) and is not coincident with the dust-enshrouded starburst
activity responsible for the far-IR emission.

To test the likelihood of LBGs being similar to these type of extremely
dusty starbursts, we randomly assigned log IRX values in the 2--3.5
range to each LBG (hashed region of Figure~2) and followed the same
aforementioned procedure to predict the X-ray luminosities. Our method
predicts the mean X-ray luminosity for the sample to be $2.7 \times
10^{43}\,$erg$\,$s$^{-1}$ (Figure~3: Panel~B). This is two orders of
magnitude larger than the observed value and strongly suggests that LBGs
are not hiding a significant amount of star formation behind a veil of
dust. In fact, $\sim92$\% of the sample would have X-ray fluxes greater
than the single source soft-band detection limit, with half being at
least an order of magnitude larger.

Furthermore, if LBGs truly have ULIG-like IRX values then the
observed far-UV fluxes imply FIR luminosities in the
$10^{12.4}$--$10^{14.7}$ L$_{\odot}$ range. Using the method described
by Adelberger \& Steidel (2000) we estimate that 96\% of the sample
would have rest-frame $\lambda =200$ \micron\ flux densities greater
than the 2 mJy detection limit for the HDF-N sub-millimeter (sub-mm)
survey conducted by Hughes et al. (1998). LBGs, however, have so far
proven to be elusive to sub-mm observers (Blain et al. 2002; Chapman et
al. 2000; Hughes et al, 1998). The recent sub-mm {\it stacked} analysis
of LBGs by Webb et al. (2002) provides a mean sub-mm $2\sigma$ detection
of only 0.414 mJy.

\subsection{Case III: Estimating X-rays Assuming SMC-type Dust}

If LBGs consist of early generation stellar populations, then low
metallicity (SMC-type) dust may be appropriate for modeling their global
far-IR emission. This would be consistent with the metallicity of the
lensed LBG MS 1512+36-cB58 ($Z \sim 1/4 Z_{\odot}$) as constrained by
Pettini et al. (2000). We test this idea is with the radiative transfer
dust models of Witt \& Gordon (2000). We compute the expected IRX and UV
spectral slope from Starburst99 (Leitherer et al. 1999) synthetic
starburst spectrum after applying extinction effects based on two Witt \&
Gordon models for SMC-dust in a shell geometry. The first model assumes
a homogeneous dust distribution in the surrounding dust shell while the
second involves a non-homogeneous 'clumpy' distribution.

The adopted Starburst99 spectrum is chosen to be consistent with the
low metallicity assumption. It is parameterized by a continuous star
formation rate, a Salpeter IMF with slope $\alpha=-2.35$, an upper
mass limit of 100 M$_{\odot}$, metallicity of $Z=0.004$, and a burst
age of 100 Myr. The bolometric dust luminosity is computed as the sum
of the energy from extincted non-ionizing photons plus the energy
deposited from {\it all} Ly$\alpha$ photons. We estimate the
Ly$\alpha$ production, prior to any extinction, by assuming that each
ionizing photon generates one Ly$\alpha$ photon (Osterbrock 1989).
The luminosity in the FIR bandpass is estimated with BC$_{\rm dust}$.

The resulting UV reddening relation for the homogeneous dust model is
similar in shape to the IRX-$\beta$ relation of MHC99, but has a
slightly less steep slope and is offset to lower values of IRX for a
values of $\beta > -2$ (dot-dash line of Figure~2). The UV reddening
relation from the alternative clumpy dust distribution is also
qualitatively similar to MHC99 but, compared to the homogeneous model,
rises faster in IRX before it turns over near $\beta \approx -1.2$ and
obtains a similar shallow slope (dashed line in Figure~2). For LBGs
in our sample with $\beta > -1.2$, the clumpy model predicts larger
values of IRX than the IRX-$\beta$ relation of MHC99. 

Having defined the UV reddening relation expected for the dust models,
we again use the same technique to predict the X-ray luminosities of the
LBG sample. The homogeneous model slightly under predicts the mean X-ray
luminosity ($1.1 \times 10^{41}\,$erg$\,$s$^{-1}$) compared to the
observed value. Within strict error limits (model and observed) the
homogeneous model agrees with the observation. The clumpy model slightly
over predicts it ($4.7 \times 10^{41}\,$erg$\,$s$^{-1}$), but the
observed mean is well within the $1\sigma$ error. The histogram in
Panel~C (Panel~D) of Figure~3 represents the homogeneous (clumpy)
model. To within the uncertainties in our technique, we are unable to
differentiate between the empirically derived relation of MHC99 and
either of the SMC-dust/shell geometry models.

Witt \& Gordon claim that the clumpy model reproduces the effective
starburst attenuation curve of Calzetti (1997; Calzetti et
al. 2000). We have calculated the UV reddening relation for the
Calzetti attenuation curve (dotted line of Figure~2) using the same
method and find it very similar to the clumpy model over the narrow
range of $\beta$ measured for the LBGs. When we use the Calzetti
effective attenuation to predict X-ray emission, we find the mean
luminosity ($4.5 \times 10^{41}\,$erg$\,$s$^{-1}$) is essentially the
same as that from the clumpy model and also in agreement with the
observed value (Figure~3: Panel~E).

We should note the effect our choice of intrinsic starburst model has
on the computed UV reddening relations. Choosing a higher metallicity
(i.e. $Z_{\odot}$) stellar population will produce a redder intrinsic
(unextincted) spectral slope ($\Delta\beta_0 \sim 0.1$) but would have
little effect on IRX values for $\beta > -2$.  Differences in either
burst age or the IMF upper mass limit will effect both $\beta_0$ and
IRX in the sense that older bursts or smaller upper mass limits will
redden $\beta_0$ and lower IRX. This would then result in a smaller
predicted X-ray flux for the LBG sample.

\subsection{Case IV: Estimating X-rays Assuming No Dust Extinction}

We can use the machinery we have developed to test the extreme
hypothesis that LBGs suffer no far-UV extinction. If the LBG sample is
actively forming stars and harbors little or no dust, we can reasonably
assume that the far-UV flux is unattenuated and dominates the bolometric
luminosity. The reddened UV colors must then be interpreted as a simple
effect of burst age. Although continuous star formation modes could not
easily account for the range of $\beta$ measured in the LBG sample,
instantaneous burst modes with ages between $30 > t_{\rm burst} > 100$
Myr can naturally redden enough to explain the observed UV properties
(Leitherer et al. 1999). However, under this scenario, the mean X-ray
luminosity is under predicted by a factor of 6 ($4.6 \times
10^{40}\,$erg$\,$s$^{-1}$). Given the large observational and model
uncertainties the difference is $\sim2\sigma$ (Figure~3: Panel~F).

\section{Conclusions}

Evidence is beginning to emerge that 2--8 keV X-rays are a good star
formation rate indicator (Ranalli et al. 2002). In local starbursts, the
2--8 keV X-ray emission is believed to be produced primarily by
high-mass X-ray binaries (HMXB; Persic \& Rephaeli 2002). This suggests
that the Brandt et al. (2001) stacking technique applied to the HDF-N
LBG sample may be detecting binary stars at $z \sim 3$. Low luminosity
AGN and the class of ultraluminous X-ray sources (ULXs or IXOs; Colbert
\& Mushotzky 1999), which may be the beamed emission from HMXBs (i.e.,
Roberts et al. 2002) or a new class of intermediate mass ($10^2$--$10^4$
M$_{\odot}$) black holes, may also contribute to the 2--8 keV
flux. Furthermore, 2--8 keV X-rays suffer little intrinsic absorption
and can escape regions where the far-UV tracers of star formation may be
heavily extincted by a dusty interstellar medium. For these reasons, the
2--8 keV X-ray luminosity strongly correlates with the bolometric
luminosity of star forming galaxies. This means that 2--8 keV X-rays can
serve as a proxy for the far-IR thermal dust emission at high-$z$.

We have derived the bolometric to 2--8 keV X-ray correlation for a local
sample of normal and starburst galaxies and have used it, in combination
with several UV reddening schemes, to predict the mean 2--8 keV X-ray
luminosity for a sample of 24 spectroscopically confirmed high redshift
Lyman-break galaxies. This simple analysis demonstrates that LBGs can not
have far-IR to far-UV flux ratios similar to those found for nearby
ULIGs, nor are they likely to be unattenuated by dust. Of the extinction
methods considered, we find that the IRX-$\beta$ starburst reddening
relation of MHC99 is the most accurate predictor of the mean X-ray
luminosity for the sample. The very similar reddening relations derived
from Witt \& Gordon (2000) extinction models of low metallicity dust in
a shell geometry and the Calzetti et al. (2000) effective starburst
attenuation curve are also consistent with the observed X-ray emission.

These results provide additional evidence that LBGs can be considered as
scaled-up local starbursts. Equally important, it suggests that
IRX-$\beta$ may be a reasonable tool for estimating the UV extinction of
high redshift LBGs. If this is the case, all 24 LBGs in this sample have
$A_{1600} < 3.1$ Mag with a mean (median) of 1.4 (1.5) Mag implying a
mean 1600\AA\ dust correction factor of $\sim4$.  This moderate level of
UV extinction is consistent with the results of Papovich et al. (2001)
who find typical 1700\AA\ correction factors of 3--4.4 from an analysis
of the UV-optical spectral energy distributions for a sample of 33 HDF-N
LBGs which includes 23 from our sample. Similar UV extinctions are also
deduced for larger LBG samples by Steidel et al. (1999) and MHC99. Our
results are the first to use low extinction X-ray emission to test and
confirm these claims.

Although it is tempting to use the results developed here to predict the
X-ray fluxes of individual LBGs, we caution the reader that our results
are statistical. The accuracy of the estimated X-ray emission for any
single LBG in this sample is only $\sim0.48$ dex. With this in mind, it
is interesting to note that when the UV reddening relations of MHC99,
Witt \& Gordon (clumpy model), and Calzetti are applied to this
technique, more than 80\% of the 2--8 keV flux originates from the half
of the sample (12) with values of $\beta > -1.5$. This may be useful for
those planning further stacking analyses as the CDF-N survey is extended
to 2 Ms.

{\em Acknowledgments}. This research has made use of the NASA/IPAC
Extragalactic Database (NED) which is operated by the Jet Propulsion
Laboratory, California Institute of Technology, under contract with the
National Aeronautics and Space Administration. This work is supported by
the NASA ADP program under the grants NAG5-8279 and NAG5-6400.

\begin{deluxetable}{lcccrrr}
\tabletypesize{\scriptsize}
\tablenum{1}
\tablecaption{Lyman-Break Galaxy Sample}
\tablewidth{0pt}
\tablehead{\colhead{HDF ID} & \colhead{RA} &
\colhead{DEC} & \colhead{z} & \colhead{f$_{8140,{\rm obs}}$} & 
\colhead{f$_{1600,{\rm rest}}$}  & \colhead{$\beta$}\\
\colhead{} & \colhead{(mm:ss.s)} &
\colhead{(mm:ss.s)} & \colhead{} & \colhead{($\mu$Jy)} & 
\colhead{($\mu$Jy)}  & \colhead{}\\
\colhead{(1)} & \colhead{(2)} &
\colhead{(3)} & \colhead{(4)} & \colhead{(5)} & 
\colhead{(6)}  & \colhead{(7)}}
\startdata
   4-858.0 &36:41.25 &12:03.1 & 3.220 & 867.04 & 757.84 & -1.43\\
   4-639.0 &36:41.71 &12:38.8 & 2.591 & 787.12 & 543.27 & -0.76\\
    2-82.1 &36:44.07 &14:09.9 & 2.267 & 459.37 & 481.26 & -2.11\\
    1-54.0 &36:44.12 &13:10.9 & 2.929 & 816.28 & 630.00 & -0.96\\
   4-445.0 &36:44.64 &12:27.4 & 2.500 &1154.41 & 812.16 & -0.81\\
   4-316.0 &36:45.09 &12:50.8 & 2.801 & 811.71 & 729.93 & -1.58\\
   2-76.11 &36:45.35 &13:46.9 & 3.160 & 262.40 & 250.39 & -1.90\\
   4-289.0 &36:46.95 &12:26.1 & 2.969 & 296.98 & 311.44 & -2.22\\
    4-52.0 &36:47.72 &12:55.8 & 2.931 & 858.38 & 622.22 & -0.70\\
   2-449.0 &36:48.34 &14:16.6 & 2.005 &1557.69 &1056.20 & -1.19\\
   4-363.0 &36:48.30 &11:45.8 & 2.980 & 345.97 & 315.61 & -1.65\\
   3-243.0 &36:49.81 &12:48.8 & 3.233 & 264.29 & 228.68 & -1.37\\
   2-525.0 &36:50.12 &14:01.0 & 2.237 & 478.23 & 368.91 & -1.48\\
   2-565.1 &36:51.19 &13:48.8 & 3.162 & 186.71 & 162.94 & -1.44\\
   2-604.0 &36:52.45 &13:37.8 & 3.430 & 398.36 & 348.52 & -1.31\\
   2-637.0 &36:52.76 &13:39.1 & 3.369 & 341.51 & 312.94 & -1.70\\
   2-834.2 &36:52.99 &14:08.4 & 3.367 &  56.69 &  55.84 & -2.19\\
   2-591.2 &36:53.18 &13:22.8 & 2.489 & 385.16 & 324.58 & -1.60\\
   2-643.0 &36:53.42 &13:29.5 & 2.991 & 514.42 & 433.17 & -1.30\\
   2-901.0 &36:53.60 &14:10.2 & 3.181 & 522.11 & 440.26 & -1.24\\
   2-824.0 &36:54.63 &13:41.4 & 2.419 & 278.61 & 281.95 & -2.06\\
   2-903.0 &36:55.08 &13:47.0 & 2.233 & 541.35 & 496.64 & -1.88\\
   3-875.0 &37:00.13 &12:25.2 & 2.050 &1079.74 & 825.69 & -1.58\\
   4-491.1 &36:43.25 &12:38.9 & 2.442 & 302.22 & 294.43 & -1.99\\
\enddata

\tablecomments{(1) ID refers to HDF-N (V2) catalog of Williams et
  al. (1996). (2) RA (J2000) plus 12 hours. (3) DEC (J2000) plus +62
  degrees. (4) Redshifts from Cohen et al. (2000) and Dawson et
  al. (2001). (5) Observed 8140\AA\ flux density from catalog of
  Williams et al. (1996). (6) Interpolated 1600\AA\ rest-frame flux
  density. (7) Photometrically derived UV spectral slope.}

\end{deluxetable}


\begin{thebibliography}{}

\bibitem[Adelberger \& Steidel(2000)]{2000ApJ...544..218A} Adelberger,
K.~L.~\& Steidel, C.~C.\ 2000, \apj, 544, 218

\bibitem[Arnaud(1996)]{1996adass...5...17A} Arnaud, K.~A.\ 1996, in ASP
Conf.~Ser.~101: Astronomical Data Analysis Software and Systems V, 5, 17

\bibitem[Blain et al.(2002)]{2002astro.ph..2228B} Blain, A.~W., Smail,
I., Ivison, R.~J., Kneib, J.-P., \& Frayer, D.~T.\ 2002, in {\it Physics
Reports}, in press, (astro-ph/0202228)

\bibitem[Brandt et al.(2001)]{2001ApJ...558L...5B} Brandt, W.~N.,
Hornschemeier, A.~E., Schneider, D.~P., Alexander, D.~M., Bauer,
F.~E., Garmire, G.~P., \& Vignali, C.\ 2001, \apjl, 558, L5

\bibitem[Brandt et al.(2001b)]{2001AJ....122.2810B} Brandt, W.~N.~et
al.\ 2001b, \aj, 122, 2810.

\bibitem[Calzetti(1997)]{1997AJ....113..162C} Calzetti, D.\ 1997, \aj,
113, 162

\bibitem[Calzetti et al.(2000)]{2000ApJ...533..682C} Calzetti, D.,
Armus, L., Bohlin, R.~C., Kinney, A.~L., Koornneef, J., \&
Storchi-Bergmann, T.\ 2000, \apj, 533, 682

\bibitem[Chapman et al.(2000)]{2000MNRAS.319..318C} Chapman, S.~C.~et
al.\ 2000, \mnras, 319, 318

\bibitem[Cohen et al.(2000)]{2000ApJ...538...29C} Cohen, J.~G., Hogg,
D.~W., Blandford, R., Cowie, L.~L., Hu, E., Songaila, A., Shopbell, P.,
\& Richberg, K.\ 2000, \apj, 538, 29

\bibitem[Colbert \& Mushotzky(1999)]{1999ApJ...519...89C} Colbert,
E.~J.~M.~\& Mushotzky, R.~F.\ 1999, \apj, 519, 89

\bibitem[Dawson et al.(2001)]{2001AJ....122..598D} Dawson, S., Stern, D., 
Bunker, A.~J., Spinrad, H., \& Dey, A.\ 2001, \aj, 122, 598. 

\bibitem[David, Jones, \& Forman(1992)]{1992ApJ...388...82D} David,
L.~P., Jones, C., \& Forman, W.\ 1992, \apj, 388, 82

\bibitem[Draine \& Lee(1984)]{1984ApJ...285...89D} Draine, B.~T.~\& Lee,
H.~M.\ 1984, \apj, 285, 89

\bibitem[Green, Anderson, \& Ward(1992)]{1992MNRAS.254...30G} Green,
P.~J., Anderson, S.~F., \& Ward, M.~J.\ 1992, \mnras, 254, 30

\bibitem[Goldader et al.(2002)]{} Goldader, J.~D., Meurer, G.~R.,
  Heckman, T.~M., Seibert, M., Sanders, D.~B., Calzetti, D., \& Steidel,
  C.~C.\ 2002, \apj, in press

\bibitem[Heckman et al.(1998)]{1998ApJ...503..646H} Heckman, T.~M.,
Robert, C., Leitherer, C., Garnett, D.~R., \& van der Rydt, F.\ 1998,
\apj, 503, 646

\bibitem[Helou, Khan, Malek, \& Boehmer(1988)]{1988ApJS...68..151H}
Helou, G., Khan, I.~R., Malek, L., \& Boehmer, L.\ 1988, \apjs, 68,
151

\bibitem[Hughes et al.(1998)]{1998Natur.394..241H} Hughes, D.~H.~et al.\
1998, \nat, 394, 241

\bibitem[Lanzetta et al.(2001)]{2001astro.ph.11129L} Lanzetta, K.~M.,
Yahata, N., Pascarelle, S., Chen, H., \& Fernandez-Soto, A.\ 2002,
\apj, in press (astro-ph/0111129)

\bibitem[Leitherer et al.(1999)]{1999ApJS..123....3L} Leitherer, C.~et
al.\ 1999, \apjs, 123, 3

\bibitem[Madau et al.(1996)]{1996MNRAS.283.1388M} Madau, P., Ferguson,
H.~C., Dickinson, M.~E., Giavalisco, M., Steidel, C.~C., \& Fruchter,
A.\ 1996, \mnras, 283, 1388

\bibitem[Marcum et al.(2001)]{2001ApJS..132..129M} Marcum, P.~M.~et al.\
2001, \apjs, 132, 129

\bibitem[Meurer et al.(1997)]{1997AJ....114...54M} Meurer, G.~R.,
Heckman, T.~M., Lehnert, M.~D., Leitherer, C., \& Lowenthal, J.\ 1997,
\aj, 114, 54

\bibitem[Meurer, Heckman, \& Calzetti(1999)]{1999ApJ...521...64M}
Meurer, G.~R., Heckman, T.~M., \& Calzetti, D.\ 1999, \apj, 521, 64 (MHC99)

\bibitem[Morrison \& McCammon(1983)]{1983ApJ...270..119M} Morrison,
R.~\& McCammon, D.\ 1983, \apj, 270, 119

\bibitem[Osterbrock(1989)]{1989agna.book.....O} Osterbrock, D.~E.\ 1989,
Astrophysics of Gaseous Nebulae and Active Galactic Nuclei, (Mill
Valley: University Science Books)

\bibitem[Papovich, Dickinson, \& Ferguson(2001)]{2001ApJ...559..620P}
Papovich, C., Dickinson, M., \& Ferguson, H.~C.\ 2001, \apj, 559, 620

\bibitem[Persic \& Rephaeli(2002)]{2002A&A...382..843P} Persic, M.~\&
Rephaeli, Y.\ 2002, \aap, 382, 843

\bibitem[Pettini et al.(2000)]{2000ApJ...528...96P} Pettini, M.,
Steidel, C.~C., Adelberger, K.~L., Dickinson, M., \& Giavalisco, M.\
2000, \apj, 528, 96

\bibitem[Rifatto, Longo, \& Capaccioli(1995)]{1995A&AS..114..527R}
Rifatto, A., Longo, G., \& Capaccioli, M.\ 1995, \aaps, 114, 527

\bibitem[Roberts et al.(2002)]{2002astro.ph..2017R} Roberts, T.~P.,
Goad, M.~R., Ward, M.~J., Warwick, R.~S., \& Lira, P.\ 2002, in {\it
New Visions of the X-ray Universe in the XMM-Newton and Chandra Era},
(The Netherlands: ESTEC), in press (astro-ph/0202017)

\bibitem[Sandage \& Tammann(1975)]{1975ApJ...196..313S} Sandage, A.~\&
Tammann, G.~A.\ 1975, \apj, 196, 313

\bibitem[Sandage \& Tammann(1981)]{1981RSA...C...0000S} Sandage, A.~\&
Tammann, G.~A.\ 1981, Revised Shapley-Ames Catalog of Bright Galaxies,
(Washington: Carnegie Inst.~of Washington)

\bibitem[Soifer et al.(1987)]{1987ApJ...320..238S} Soifer, B.~T.,
Sanders, D.~B., Madore, B.~F., Neugebauer, G., Danielson, G.~E., Elias,
J.~H., Lonsdale, C.~J., \& Rice, W.~L.\ 1987, \apj, 320, 238

\bibitem[Steidel et al.(1999)]{1999ApJ...519....1S} Steidel, C.~C.,
Adelberger, K.~L., Giavalisco, M., Dickinson, M., \& Pettini, M.\ 1999,
\apj, 519, 1

\bibitem[Webb et al.(2002)]{2002astro.ph..1181W} Webb, T.~M.~A.~et
al.\ 2002, \apj, in press (astro-ph/0201181)

\bibitem[Williams et al.(1996)]{1996AJ....112.1335W} Williams,
R.~E.~et al.\ 1996, \aj, 112, 1335

\bibitem[Witt \& Gordon(2000)]{2000ApJ...528..799W} Witt, A.~N.~\&
Gordon, K.~D.\ 2000, \apj, 528, 799


\end{thebibliography}
\end{document}